\documentclass[11pt]{article}

\usepackage{amsmath,amssymb,amsthm,bm}
\usepackage[dvipdfmx]{graphicx}
\usepackage{psfrag,epsf}
\usepackage{enumerate}

\usepackage{natbib}
\newcommand{\blind}{0}

\usepackage{multicol}
\usepackage{flafter}
\usepackage{color,hyperref,xcolor}
\hypersetup{colorlinks=true,urlcolor=blue,citecolor=purple}
\usepackage{cleveref}
\numberwithin{equation}{section}

\def\R{\mathbb{R}}
\def\I{\mathbf{I}}

\def\Y{\mathbf{Y}}
\def\N{\mathbb{N}}
\def\bmu{\bm{\mu}}

\newcommand{\e}{\mathbf{e}}
\newcommand{\w}{\mathbf{w}}
\newcommand{\abs}[1]{\left\lvert #1 \right\rvert}
\newcommand{\norm}[1]{\left\| #1 \right\|}
\newcommand{\cov}{\mathrm{Cov}}
\newcommand{\eps}{\varepsilon}
\newcommand{\given}{\hspace{2pt}\vert\hspace{2pt}}
\newcommand{\dash}{^{\prime}}

\newcommand{\poi}{\mathrm{Poisson}}

\DeclareMathOperator*{\argmin}{arg\,min}
\newcommand{\ind}{\mathbb{I}}

\newtheoremstyle{general}
{3mm} 
{3mm} 
{} 
{} 
{\bfseries} 
{.} 
{.5em} 
{} 

\theoremstyle{general}
\newtheorem{definition}{Definition}

\begin{document}

\def\spacingset#1{\renewcommand{\baselinestretch}%
{#1}\small\normalsize} \spacingset{1}


\if0\blind
{
  \title{\bf Analyzing count data using a time series model with an exponentially decaying covariance structure}
  \author{Soudeep Deb\thanks{Corresponding author. Email: soudeep@iimb.ac.in. ORCiD: 0000-0003-0567-7339}\hspace{.2cm}\\
    Indian Institute of Management, Bangalore \\ Bannerghatta Main Rd, Bilekahalli \\ Bengaluru, Karnataka 560076, India.}
  \maketitle
} \fi

\if1\blind
{
  \bigskip
  \bigskip
  \bigskip
  \begin{center}
    {\Large\bf Analyzing count data using a time series model with an exponentially decaying covariance structure}
\end{center}
  \medskip
} \fi

\bigskip
\begin{abstract}
Count data appears in various disciplines. In this work, a new method to analyze time series count data has been proposed. The method assumes exponentially decaying covariance structure, a special class of the Mat\'ern covariance function, for the latent variable in a Poisson regression model. It is implemented in a Bayesian framework, with the help of Gibbs sampling and ARMS sampling techniques. The proposed approach provides reliable estimates for the covariate effects and estimates the extent of variability explained by the temporally dependent process and the white noise process. The method is flexible, allows irregular spaced data, and can be extended naturally to bigger datasets. The Bayesian implementation helps us to compute the posterior predictive distribution and hence is more appropriate and attractive for count data forecasting problems. Two real life applications of different flavors are included in the paper. These two examples and a short simulation study establish that the proposed approach has good inferential and predictive abilities and performs better than the other competing models.
\end{abstract}

\noindent%
\vfill

\newpage

\section{Introduction}
\label{sec:introduction}

Integer-valued time series or count data appears in many disciplines, ranging from economics to public health to social sciences. Popular examples of such data are the number of people affected from a virus, the number of a certain product sold per day, the number of website visits, the number of extreme environmental events at a location or the number of accidents at an intersection. Generalized linear models (GLM) with Poisson or negative binomial distribution are suitable to deal with the discreteness (see \cite{mccullagh2018generalized} for an in-depth discussion on GLM) and they can assess the effect of different regressors on the response variable, but they fail to address the correlated nature of the data. Meanwhile, models like autoregressive integrated moving average (ARIMA) can analyze the covariance structure for a real-valued time series in an appropriate way, but as \cite{freeland2004forecasting} pointed out, they are also inappropriate for count data as they do not produce coherent forecasts. In fact, modelling count data demands one to consider both the discreteness and the time-dependence properties of the series. \citet{mckenzie1985some} is first of the many significant papers in this regard. He proposed binomial thinning, which paved the way to the class of integer-valued autoregressive (INAR) and integer-valued moving average (INMA) processes. As the name of the model suggests, these are extensions of standard real-valued AR and MA models. \cite{quddus2008time} and \cite{weiss2008serial} are noteworthy for related discussions and applications.  This family of models are further generalized to include integer valued autoregressive moving average (INARMA) processes in presence of exogenous regressors, cf. \cite{neal2007mcmc} and \cite{enciso2009efficient}. A similar flavor in the modeling approach can be found in discrete autoregressive moving average (DARMA) models and generalized linear autoregressive moving average (GLARMA) models. For relevant readings, refer to \cite{chang1984daily}, \cite{dunsmuir2015generalized} and the references therein. Autoregressive conditional Poisson (ACP) model is another attractive approach which deals with the issues of discreteness, potential over-dispersion (when the variance is greater than the mean) and serial correlation. It was first introduced in detail by \cite{heinen2003modelling}. \cite{ghahramani2009some} discussed some interesting properties of this model while \cite{fokianos2009poisson} laid out an extensive theoretical background on Poisson autoregression. A related application can be found in \cite{gross2013predicting}. 

Each of the aforementioned papers has its own advantage and contribution. However, the effectiveness of different approaches lie with successfully capturing the underlying dependence pattern of the data. To that end, we present an attractive framework for the covariance structure of count data. We develop a hierarchical process using a latent variable that governs the Poisson distributed count data in a non-linear fashion. This latent variable is further assumed to be an additive combination of some fixed effects, a white noise and a correlated process equipped with a Mat\'ern class of covariance function. While somewhat similar models have been adopted in this regard, they come with severe complexities both in terms of estimation and in terms of prediction. Our approach, on the contrary, uses a Bayesian framework with appropriate Gibbs sampling and adaptive rejection metropolis sampling algorithms, and thereby obviates the aforementioned complexity and makes the numerical analysis easier. Consequently, the proposed method enables better forecasting accuracy than the existing models relevant to the data. Further, the specification of the model does not require the time points to be consecutive and allows an irregular index set. This in turn helps one to efficiently deal with any type of discrete time series data.

In what follows, first we shall discuss the model, along with the Bayesian framework and implementation procedure, in \Cref{sec:methods}. A short simulation study is presented in \Cref{sec:simulation}. The following section is about real life application of the proposed method. Two different types of dataset are considered in this work and they are discussed in \Cref{subsec:accident} and \Cref{subsec:bikes}. We will conclude the paper with some important remarks in \Cref{sec:remarks}. 

\section{Methods}
\label{sec:methods}

\subsection{Model}
\label{subsec:model}

Let $t_i$, for $i\in\{1,2,\hdots,T\}$, denote a time point. It is worth mention that our method is flexible enough to allow any temporal resolution, and hence $t_i$ can represent hour, day, week etc, depending on the data. As indicated earlier, the time points can be irregularly spaced. We use $\Gamma$ to denote the index set $\{t_1,t_2,\hdots,t_T\}$. 

Throughout the following discussion, any boldfaced letter $\bm{b}$ denotes a vector of the form $(b_{t_1},b_{t_2},\hdots,b_{t_T})$, while $\I_{n}$ denotes an identity matrix of order $n\times n$. For a set $A$, $\abs{A}$ is the cardinality. Also, whenever used, $\N$ and $\R$ denote the set of natural numbers and the set of real numbers, respectively.

For each $t\in\Gamma$, let $Y_t$ be the value of the response variable. In vectorized form, $\Y$ denotes the vector of all outcomes $(Y_{t_1,}Y_{t_2},\hdots,Y_{t_T})$. We assume $Y_t$ to have a Poisson distribution with parameter $\lambda_t$. In our modeling scheme, we use the log-link for the Poisson parameter. Let $\mu_t = \log \lambda_t$. Corresponding to $Y_t$, $X_t$ is the column vector for all covariates used in the mean structure of the model. $X$ is the overall design matrix such that $X\dash=[X_{t_1}:X_{t_2}:\hdots:X_{t_T}]$.

Our proposed model starts with the aforementioned assumption that for a latent variable $\mu_t$, $Y_t \sim \poi(e^{\mu_t})$. $\mu_t$ depends on the covariates through an additive structure. We assume that, given the values of the covariates, the outcomes are temporally dependent of each other. In that light, we consider a hierarchical structure for $\mu_t$ as follows. 
\begin{equation}
\label{eqn:hierarchical-model}
\mu_t = X_t\dash \beta + \eps_t, \quad \eps_t = w_t + e_t.
\end{equation}

Here, $\beta$ is a parameter vector of appropriate order, and the term $X_t\dash \beta$ captures the effects of all seasonal, continuous or discrete type covariates on the mean structure. $(\eps_t)_{t\in\Gamma}$ is a time dependent error process and is an additive combination of two terms. $(w_t)_{t\in\Gamma}$ denotes a zero-mean temporally correlated process and $(e_t)_{t\in\Gamma}$ stands for a zero-mean independent and identically distributed (iid) white noise process. Mathematically, assume that $e_t$'s are iid $N(0,\sigma^2)$. For $w_t$, we take a correlation structure that decays exponentially. In particular, the covariance between $w_{t_1}$ and $w_{t_2}$ is
\begin{equation}
\label{eqn:covariance}
\cov(w_{t_1},w_{t_2}) = \sigma_w^2 \ \exp(-\phi d(t_1,t_2)). 
\end{equation}

$d(\cdot,\cdot)$ is a distance function and is defined as the absolute difference between the two time points. Note that this covariance function is a special choice ($\nu=0.5$) among the Mat\'ern class of covariance functions, cf. \cite{minasny2005matern}. Hereafter, for convenience, let us denote the covariance matrix of $\w = (w_{t_1},w_{t_2},\hdots,w_{t_T})$ by $\sigma_w^2\Sigma_w$. Then, following the notations described above, the complete modeling scheme can be written as follows.
\begin{equation}
\label{eqn:full-model}
    Y_t \sim \poi\left(e^{\mu_t}\right), \ \bmu = X\beta + \w + \e, \ \w \sim N(0,\sigma_w^2\Sigma_w), \ \e \sim N(0,\sigma^2\I_{T}).
\end{equation}

\subsection{Bayesian implementation}
\label{subsec:estimation}

In order to implement the above modeling scheme, a Bayesian framework is used. For that, we need to assign prior distributions to the parameters in the study. Each components in $\beta$, a priori, is assumed to be independent with flat (uniform) prior distribution on $\R$. $\sigma^2$ and $\sigma_w^2$ are variance parameters and independent inverse gamma (IG) priors with parameters $(a, b)$ are used for them. These in fact are conditionally conjugate priors (see, for example, \cite{gelman2006prior}) for normal likelihood. We take $a>2$ to ensure finite mean and variance for the prior distribution. The parameter $\phi$ in the covariance function is considered to be fixed throughout the analysis, and we estimate it using a cross-validation scheme, details of which is provided at the end of this subsection.

Let us now look into the posterior distributions of the parameters. Throughout the discussion below, $K$ indicates a constant term, that may vary from time to time. Recall the model defined by \cref{eqn:full-model}. The log-likelihood is 
\begin{align}
    l(\beta,\sigma^2,\sigma_w^2,\bmu,\w; \Y) &= K - \sum_{t\in\Gamma} e^{\mu_t} + \sum_{t\in\Gamma}\mu_t Y_t - \frac T2\log \sigma^2 - \frac{1}{2\sigma^2}\norm{\bmu - X\beta - \w}^2 \nonumber \\
    \label{eqn:likelihood}
    & - \frac T2\log \sigma_w^2 - \frac{1}{2\sigma_w^2}\w\dash\Sigma_w^{-1}\w.
\end{align}

Then, using the prior distributions described earlier in this section, the joint posterior distribution can be written as
\begin{align}
\log \pi(\beta,\sigma^2,\sigma_w^2,\bmu,\w \given \Y) &= K - \sum_{t\in\Gamma} e^{\mu_t} + \Y\dash\bmu - \frac{1}{2\sigma^2}\norm{\bmu - X\beta - \w}^2 - \frac{1}{2\sigma_w^2}\w\dash\Sigma_w^{-1}\w \nonumber \\
\label{eqn:joint-posterior}  
& - \left(a+\frac T2+1\right)\log \sigma^2 - \left(a+\frac T2+1\right)\log \sigma_w^2 - \frac{b}{\sigma^2} - \frac{b}{\sigma_w^2}.
\end{align}

The above expression is complicated enough to find out a closed form for the joint posterior distribution. Thus, for ease of computation, the principles of Gibbs sampling are used. It is a type of Markov chain Monte Carlo (MCMC) method for obtaining a sequence of realizations from a specified multivariate probability distribution. Extensive reading on Gibbs sampling can be found in \cite{geman1984stochastic} and \cite{durbin2002simple}. In this method, every parameter is updated in an iterative way using the conditional posterior distributions given the other parameters. 

For example, in case of $\sigma^2$, the conditional posterior distribution is obtained as
\begin{equation*}
\log \pi(\sigma^2\given \Y,\bmu,\beta,\w) = K - (a+T+1)\log \sigma^2 - \frac{1}{\sigma^2}\left(b + \frac12\norm{\bmu - X\beta - \w}^2\right).
\end{equation*}

Note that the above is in fact the logarithm of an inverse gamma density function with shape parameter $a+T/2$ and scale parameter $\norm{\bmu - X\beta - \w}^2/2$. Thus,
\begin{equation}
\label{eqn:sigma-posterior}
(\sigma^2\given \Y,\bmu,\beta,\w) \sim \mathrm{IG} \left(a+\frac T2, b+\frac12\norm{\bmu - X\beta - \w}^2\right).
\end{equation}

In an exact similar way, 
\begin{equation}
\label{eqn:sigmap-posterior}
(\sigma_w^2\given \Y,\bmu,\beta,\w) \sim \mathrm{IG}\left(a+\frac T2, b+\frac{1}{2}\w\dash\Sigma_w^{-1}\w\right).
\end{equation}

For $\beta$, the conditional distribution can be written as
\begin{equation*}
\log \pi(\beta \given \Y,\bmu,\sigma^2,\w) = K - \frac12 \beta\dash \biggl(\frac{X\dash X}{\sigma^2}\biggr)\beta + \frac12 \cdot 2\beta\dash  \cdot \frac{X\dash (\bmu-\w)}{\sigma^2}.
\end{equation*}

It is clearly the log-density of a multivariate normal distribution. Straightforward algebraic manipulations then imply that the conditional posterior of $\beta$ is 
\begin{equation}
\label{eqn:theta-posterior}
(\beta \given \Y,\bmu,\sigma^2,\w) \sim N\left((X\dash X)^{-1}X\dash (\bmu-\w),\sigma^2(X\dash X)^{-1}\right).
\end{equation}

Using identical techniques, we can get the posterior distribution for $\w$ as
\begin{equation}
\label{eqn:w-posterior}
(\w \given \Y,\bmu,\beta,\sigma^2,\sigma_w^2) \sim N\left((\I_T+\frac{\sigma^2}{\sigma_w^2}\Sigma_{w}^{-1})^{-1}(\bmu-X\beta), (\frac{1}{\sigma^2}\I_T+\frac{1}{\sigma_w^2}\Sigma_{w}^{-1})^{-1}\right).\end{equation}

Finally, we find the posterior distribution of $\bmu$. Note that conditional on data and other parameters, each component of $\bmu$ is independent of others. The corresponding posterior is
\begin{equation}
\label{eqn:mu-posterior}
    \log \pi(\mu_t\given Y_t,\beta,w_t) = K - \frac{1}{2\sigma^2}\left(\mu_t - (X_t\dash\beta + w_t + \sigma^2Y_t)\right)^2 - e^{\mu_t}.
\end{equation}

Simulating directly from the above posterior is difficult, since the normalizing constant or the distribution function cannot be obtained in a straightforward way. Hence, at this stage, we use adaptive rejection metropolis sampling (ARMS) algorithm. It is a method for efficiently sampling from complicated univariate densities, as discussed in \cite{gilks1992adaptive} and \cite{gilks1995adaptive}.

The above conditional posterior distributions are iterated multiple times and existing theory ensures that the algorithm will converge to the true posterior distributions. In the process, the Gibbs sampling generates a Markov chain of posterior samples and we shall use that to make inference. To that end, a crucial point to remember is that each of the samples is correlated with nearby realizations. Also, samples from the beginning of the chain may not represent the true posterior distribution. Naturally, appropriate care should be taken both to ensure convergence of the chains and to generate independent samples from the posterior distributions. For monitoring the convergence of the sampler, we use the Gelman-Rubin statistic, cf. \cite{gelman1992inference}. In this approach, multiple parallel chains are considered with different starting values. Since they all eventually converge to the stationary distribution, after enough iterations, it should be impossible to distinguish between the chains. The Gelman-Rubin statistic leverages this idea and compares the variation between the chains to the variation within the chains. Ideally, the statistic approaches the value of 1. While implementing the Gibbs sampler in our method, we start multiple chains by randomly generating initial values and monitor the value of the statistic until it is below 1.5 (hereafter denoted as the burn period). Then we take a large posterior sample sufficiently apart from each other so as to ensure independence. 

Finally, as mentioned before, the parameter $\phi$ (refer to \cref{eqn:covariance}) is unknown and in order to find out the optimal value, a cross-validation (CV) scheme is used. This technique has been used in some other similar modeling works, for example \cite{sahu2006spatio} and \cite{deb2019spatio}. In our case, depending on the temporal resolution of the data, a range of values for $\phi$ is chosen. Let us call it $S$. Next, we set aside a validation set ($V$), and train the model on the rest of the data for different values of $\phi$. The predictive performance of each case is then evaluated based on an appropriate scoring rule (see the following two subsections for more details). Formally, for a model trained with $\phi\in S$, let $\hat{f}_{\phi,t}$ denote the posterior predictive distribution for true $Y_t$ at time point $t\in V$. Let $L(Y_t,\hat f_{\phi,t})$ denote the model evaluation criteria. The optimal choice of $\phi$ is then defined as
\begin{equation}
\label{eqn:validation-mse}
\phi_{\mathrm{opt}} = \argmin_{\phi \in S} L\left(Y_t,\hat{f}_{\phi,t}\right).
\end{equation}

Combining all of the above results, the steps of the Gibbs sampler for a fixed $\phi$ can now be presented in the following way.

\begin{itemize}
    \item[Step 1.] Fix a value of the parameter $\phi$. 
    \item[Step 2.] Take multiple initial values for $\sigma^2$, $\sigma_w^2$, $\w$, $\beta$ and start multiple chains.
    \item[Step 3.] Compute the initial value of $\bmu$ as $\bmu= X\beta$. 
    \item[Step 4.] Use current values of $\beta$, $\bmu$, $\w$ in \cref{eqn:sigma-posterior} and \cref{eqn:sigmap-posterior} to generate posterior realizations for the variance parameters.
    \item[Step 5.] Use current values of $\sigma^2$, $\bmu$, $\w$ to generate posterior realization for $\beta$, through \cref{eqn:theta-posterior}.
    \item[Step 6.] Use current values of $\sigma^2$, $\sigma_w^2$, $\beta$, $\bmu$ to generate posterior realization for $\w$, through \cref{eqn:w-posterior}.
    \item[Step 7.] Use ARMS algorithm to generate posterior realization for $\mu$ through \cref{eqn:mu-posterior}, using the current values of $\sigma^2$, $\beta$, $\w$.
    \item[Step 8.] Check for convergence of the chains using the Gelman-Rubin statistic. If convergence is attained, go to Step 9. Else, repeat steps 4 to 7.
    \item[Step 9.] Repeat the algorithm (steps 4 to 7) $M$ times more (for large $M$) and store every $m$th realizations (for $m>10$) in the posterior sample.
\end{itemize}

In all of the applications in this paper, posterior samples of size 1000 are generated. We also find out that the burn period in all applications are below 5000 iterations. All computations in this study are executed in RStudio version 1.2.5033, coupled with R version 3.6.2. 



\subsection{Future prediction}
\label{subsec:prediction}

A crucial contribution of this paper is to use the above model to provide a prediction strategy for $Y_{t\dash}$, $t\dash>T$. Let $X_t\dash$ denote the corresponding column vector of covariates. $\mu_{t\dash}$ and $w_{t\dash}$ are defined accordingly. Note that the posterior predictive distribution $f(Y_{t\dash}\given \Y)$ can be written as
\begin{equation}
    f(Y_{t\dash}\given \Y) \propto \int f(Y_{t\dash}\given \mu_{t\dash})\pi(\mu_{t\dash}\given \Y) \ d\mu_{t\dash}.
\end{equation}

Now, instead of solving the above integral, a better and more convenient idea is to use the Gibbs sampler estimates to draw observations from the posterior predictive distribution. One can do it sequentially. At first, draw samples for $\bmu,\w,\beta,\sigma^2,\sigma_w^2$ using the conditional posteriors derived above. Then, we need to simulate $w_{t\dash}$ using the conditional distribution of $(w_{t\dash}\given \w,\sigma_w^2)$. To compute it, observe that
$$\left(\begin{array}{c} \w \\ w_{t\dash}\end{array}\right) \sim N\biggl(0,\sigma_w^2\left[\begin{array}{cc} \Sigma_{w} & \Sigma_{t-t\dash}  \\ \Sigma_{t-t\dash}\dash & 1 \end{array}\right]\biggr).$$

Here, $\Sigma_{t-t\dash}$ is the column vector denoting the covariance of $w_{t\dash}$ with the elements of $\w$. Following \cref{eqn:covariance}, the $j$th element of $\Sigma_{t-t\dash}$ is $\exp(-\phi d(t_j,t\dash))$. Thus, the principle of conditional distribution for multivariate normal distribution imply that
\begin{equation}
\label{eqn:wnew-conditional}
(w_{t\dash}\given \w,\sigma_w^2) \sim N\left(\Sigma\dash_{t-t\dash}\Sigma_w^{-1}\w, \sigma_w^2\left(1-\Sigma\dash_{t-t\dash}\Sigma_w^{-1}\Sigma_{t-t\dash}\right)\right).
\end{equation}

Then, using the above realizations, an estimate for $\mu_{t\dash}$ can be obtained by generating samples from a normal distribution with mean $X_{t\dash}\dash \beta+w_{t\dash}$ and variance $\sigma^2$. Finally, posterior predictive distribution for $Y_{t\dash}$ is given by a Poisson distribution with parameter $\exp(\mu_{t\dash})$. This predictive distribution is the most significant component to judge the accuracy, and is used in different scoring rules. More discussions on that will follow in the subsequent section.  Further, if needed, the predictive distribution can be used to get a point forecast for $Y_{t\dash}$ while a prediction interval can be computed using many samples from the posterior predictive distribution. 

\subsection{Evaluation and comparison}
\label{subsec:evaluation}

In the real life applications in this paper, we would assess the performance of the proposed method and provide a comparative study against GLARMA and ACP method (discussed in \Cref{sec:introduction}), two of the most popular techniques for discrete time series data. We shall also include the results of GLM as a baseline model.

To assess the goodness of fit of the models, we shall use the standard measures such as log-likelihood, $R^2$ and root mean squared error (RMSE). However, as \cite{kolassa2016evaluating} discussed at length, usual measures of comparing point predictions with the true values, such as mean squared error (MSE), mean absolute deviation (MAD), mean absolute percentage error (MAPE) etc are not appropriate in the context of count data. One should instead evaluate the accuracy of a discrete time series model using the full predictive distribution. Probability integral transform (PIT), on that note, is a relevant technique for continuous data, but it also suffers from major drawbacks in case of count data. See \cite{gneiting2007probabilistic} for a related discussion. The authors there described how different scoring rules can address the shortcoming of the above measures and can work as useful tools for assessing both the calibration and the sharpness of the predictive distributions.

Let us use the simplified notation $\hat f_{t\dash} = (\hat f_{t\dash,k})$, for possible realizations $k\in \N+\{0\}$, to denote the predictive distribution obtained from a model for a future time point ${t\dash}$. Let $y$ be a single realization, usually the true value in the data. Then, a scoring rule is defined as a function $s$ which maps $\hat f_{t\dash}$ and $y$ to a penalty value $s(\hat f_{t\dash}, y)$. Further, the scoring rule is proper if its expected value is minimal when $\hat f_{t\dash}$ is the true future distribution of $y$. The reader is referred to \cite{czado2009predictive} for a discussion of several proper scoring rules in the context of count data. In this work, we are going to use the following three scores to evaluate and compare the competing models.

\begin{definition}
Let $\norm{\hat f_{t\dash}}=(\sum \hat f_{t\dash,k}^2)^{1/2}$. Then, the Brier score or quadratic score is defined as
\begin{equation}
\label{eq:brier}
    BS(\hat f_{t\dash},y) = -2\hat f_{t\dash,y} + \norm{\hat f_{t\dash}}^2. 
\end{equation}
\end{definition}

\begin{definition}
Using $\norm{f_{t\dash}}$ as above, the spherical score is defined as
\begin{equation}
\label{eq:spherical}
    SpS(\hat f_{t\dash},y) = -\frac{\hat f_{t\dash,y}}{\norm{\hat f_{t\dash}}}. 
\end{equation}
\end{definition}

\begin{definition}
Let $(\hat F_{t\dash,k})$ be the cumulative predictive distribution corresponding to $\hat f_{t\dash}$. Then, the ranked probability score is defined as
\begin{equation}
\label{eq:rps}
    RPS(\hat f_{t\dash},y) = \sum \left(\hat F_{t\dash,k} - \ind\{y\le k\}\right)^2. 
\end{equation}
\end{definition}

Note that the above definitions are for a single $t\dash$. In case of predicting more than one step ahead, we shall compute the mean of the scores of all future time points.

\section{Simulation study}
\label{sec:simulation}

We focus on three different toy examples in this section. Keeping in line with the model in \cref{eqn:full-model} and also with the applications in the following section in mind, we consider the following generic structure to simulate the data $\Y=(Y_t)_{t\in\Gamma}$. 
\begin{equation}
    \label{eq:simulation-model}
    Y_t \sim \poi(e^{\mu_t}), \; \mu_t = \beta_0 + \beta_1X_{1,t} + \beta_2X_{2,t} + \varepsilon_t.
\end{equation}

\begin{figure}[!ht]
\begin{center}
\includegraphics[width=0.8\textwidth,keepaspectratio]{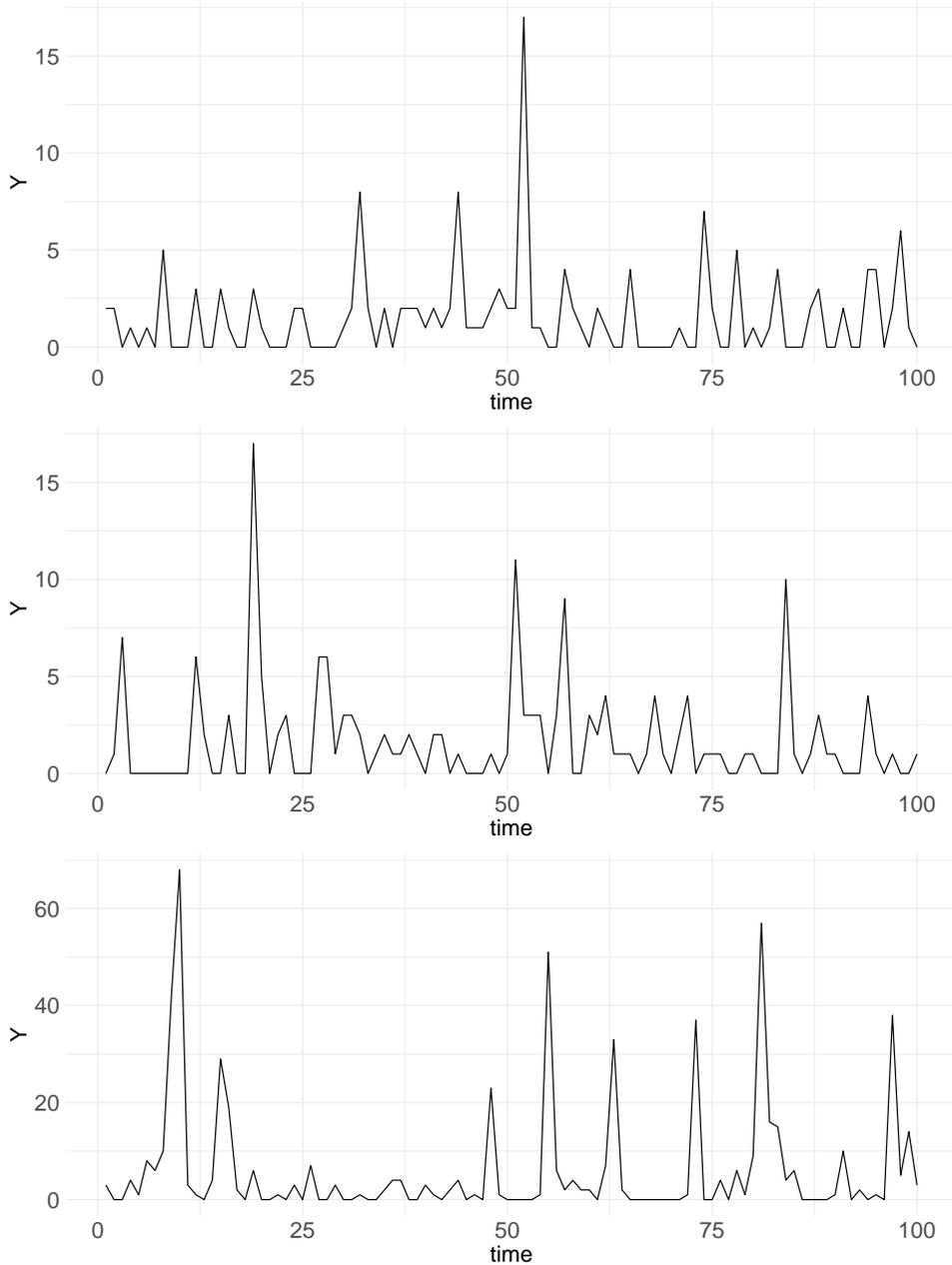}
\end{center}
\caption{Three simulated series: (Top) $\varepsilon_t$'s are iid standard normal random variables, (Middle) $\varepsilon_t$'s are simulated from an AR(3) process, (Bottom) $\varepsilon_t$'s are simulated as $w_t + e_t$, where $w_t$'s are coming from an AR(1) process and $e_t$'s are iid $t_5$.}
\label{fig:simulated-series}
\end{figure}

Here, $X_{1,t}$ and $X_{2,t}$ correspond to the values of two regressors and in all simulation experiments, these covariates are randomly generated. The coefficients are taken as $\beta_0=0.15$, $\beta_1=-0.28$ and $\beta_3=0.18$. For $(\varepsilon_t)_{t \in \Gamma}$, we consider three different data generating processes (DGP) to analyze the efficacy of the proposed approach. First, we take $\varepsilon_t$'s to be iid standard normal random variables. Second, we introduce a dependence structure in $(\varepsilon_t)_{t \in \Gamma}$, and they are simulated from an autoregressive (AR) process of order 3, along with standard normal innovations. The AR coefficients used are $(0.2,-0.3,0.1)$. Third, $\varepsilon_t$ is simulated as $w_t + e_t$, where $(w_t)_{t \in \Gamma}$ is an AR process of order 1 (with AR coefficient 0.5) and $(e_t)_{t \in \Gamma}$ are iid $t$-distributed random variables with degrees of freedom 5. In all experiments, for convenience, we use a regular index set $\Gamma$ with $\abs{\Gamma}=100$. Examples of the simulated series from the three different processes are presented in \Cref{fig:simulated-series}. Note that the third process assigns stronger autocorrelation and higher variability than the second. 

Our objectives in these experiments are two fold. On one hand, we want to find out the errors in estimating the coefficients of the effects of the regressors while on the other, we want to evaluate the prediction accuracy of different methods. At this stage, it is worth recall that while implementing our proposed method, we need to estimate $\phi$ using a CV scheme (refer to \Cref{subsec:estimation}). For that, the time series we analyze is divided into a CV train series (using approximately 90\% of the observations) and validation is done on the remaining part. Possible choices of $\phi$ are taken as $0.01,0.1,0.25,0.5,1,1.5,3$.  Experimentation showed that the results are stable around these values and so, for computational ease, we do not consider a finer grid. On the other hand, appropriate orders for both GLARMA and ACP are chosen using Akaike Information Criterion (AIC). Maximum lags used in these cases are 7, thereby ensuring both parsimony and effectiveness. R packages glarma (\cite{glarmapackage}) and acp (\cite{acppackage}) are used for the implementation of these two models. We repeat each experiment many times and find out the mean squared error (MSE) in estimating the three coefficients. These results are presented in \Cref{tab:mse-simulation}. 

\begin{table}[!htb]
\centering
\caption{Mean squared error in estimating the coefficients in the simulated experiments. DGP 1 corresponds to the iid choice, DGP 2 is the AR(3) choice and DGP 3 denotes the combination of AR(1) and iid $t_5$ distributions for $\varepsilon_t$.}
\label{tab:mse-simulation}
\begin{tabular}{lccccc}
  \hline
  DGP & Parameter & GLM & GLARMA & ACP & Our model \\ 
  \hline
  DGP 1   & $\beta_0$ & $0.2920$ & $0.3242$ & $0.3141$ & $0.3111$ \\ 
            & $\beta_1$ & $0.2685$ & $0.3015$ & $0.2643$ & $0.0622$ \\ 
            & $\beta_2$ & $0.0206$ & $0.0242$ & $0.0215$ & $0.0158$ \\
  \hline
  DGP 2   & $\beta_0$ & $0.3136$ & $0.2975$ & $0.2252$ & $0.3415$ \\ 
            & $\beta_1$ & $0.1943$ & $0.2130$ & $0.2643$ & $0.0606$ \\ 
            & $\beta_2$ & $0.0231$ & $0.0228$ & $0.0229$ & $0.0141$ \\
  \hline
  DGP 3   & $\beta_0$ & $5.5239$ & $2.4782$ & $4.5365$ & $1.0033$ \\ 
            & $\beta_1$ & $2.6937$ & $1.6937$ & $2.9678$ & $0.1287$ \\ 
            & $\beta_2$ & $0.3500$ & $0.1676$ & $0.3477$ & $0.0241$ \\
  \hline
\end{tabular}
\end{table}

We can see that the proposed approach outperforms the other three methods in estimating the coefficients across all scenarios. ACP does not observe substantial improvement from GLM in any of the three cases. For the third DGP, where we introduce the most complicated structure for $\varepsilon_t$, GLARMA turns out to be the second best, although the errors are more than double than that of our method.

In order to assess the forecasting performance, we set aside 10\% of the observations and calculate the scoring rules described in the previous section. \Cref{tab:pred-simulation} presents the prediction scores of the candidate models in the three different cases. Here, we note that all methods perform similarly in the iid case. Both in terms of Brier score and spherical score, GLM works better for the second case where $(\varepsilon_t)_{t \in \Gamma}$ are mildly autocorrelated. Finally, for the third DGP, prediction performances turn out to be the best in our approach.

\begin{table}[!htb]
\centering
\caption{Mean prediction score in the simulated experiments. DGP 1 corresponds to the iid choice, DGP 2 is the AR(3) choice and DGP 3 denotes the combination of AR(1) and iid $t_5$ distributions for $\varepsilon_t$.}
\label{tab:pred-simulation}
\begin{tabular}{llcccc}
  \hline
  DGP & Scoring rule & GLM & GLARMA & ACP & Our model \\ 
  \hline
  DGP 1   & Brier score               & $-0.1477$ & $-0.1638$ & $-0.1616$ & $-0.1515$ \\ 
            & Spherical score           & $-0.4542$ & $-0.4080$ & $-0.4059$ & $-0.3905$ \\ 
            & Ranked probability score  & $0.0001$ & $0.0001$ & $0.0001$ & $0.0001$ \\
  \hline
  DGP 2   & Brier score               & $-0.1938$ & $-0.1640$ & $-0.1549$ & $-0.1560$ \\ 
            & Spherical score           & $-0.4741$ & $-0.4073$ & $-0.3961$ & $-0.3943$ \\ 
            & Ranked probability score  & $0.0001$ & $0.0001$ & $0.0001$ & $0.0001$ \\
  \hline
  DGP 3   & Brier score               & $-0.0934$ & $0.0174$ & $0.0305$ & $-0.1104$ \\ 
            & Spherical score           & $-0.3360$ & $-0.1620$ & $-0.1222$ & $-0.3282$ \\ 
            & Ranked probability score  & $0.0006$ & $0.0109$ & $0.0112$ & $0.0006$ \\
  \hline
\end{tabular}
\end{table}

All in all, we see that the proposed approach has much better applicability in various scenarios. Especially to estimate the effect of the covariates, our method is considerably better than other candidate models. Prediction scores are also very good for our model.

\section{Application}
\label{sec:results}

\subsection{Road accidents in Britain}
\label{subsec:accident}

In the first real data application, we study the monthly number of drivers killed in Great Britain during the years of 1969 to 1984. This data was first discussed in detail by \cite{harvey1986effects} who analyzed the effect of a new legislation which came in January of 1983. That rule mandated the wearing of seat-belts in light goods vehicles. This dataset is freely available in R. Along with the information on the number of deaths of different types, some additional covariates are also provided. We shall use `kilometers' (distance driven by the cars, in thousands) and `petrol price' in this study. In the aforementioned paper, the authors considered the numbers of casualties for drivers and for passengers of the cars, the numbers of which are on the larger side. On the other hand, the monthly number of casualties for the drivers of vans is much smaller. It ranges between 2 and 17, with a mean of 9.06 and a variance of 13.2. Clearly, this time series is an attractive one for the purpose of the current paper. In the following discussion, the above variable is denoted as `casualties' for convenience. Summary of this and the two covariates are presented in \Cref{tab:summary-accident}.

\begin{table}[!htb]
\caption{Summary of the response variable and the covariates for road accident data.}
\label{tab:summary-accident}
\centering
\begin{tabular}{lcccc}
  \hline
 & Range & Mean & Median & Standard deviation \\ 
  \hline
Casualties & $(2,17)$ & 9.057 & 8 & 3.637 \\ 
Kilometers (in thousands) & $(7.685,21.626)$ & 14.994 & 14.987 & 2.938 \\ 
Petrol price & $(0.081,0.133)$ & 0.104 & 0.104 & 0.012 \\ 
   \hline
\end{tabular}
\end{table}

In this analysis,  we take the data until December 1983 as our train series and leave the last year of data as our test set. We follow the same idea as in the previous section to estimate $\phi$ in our model. Once again, we consider the choices $0.01,0.1,0.25,0.5,1,1.5,3$, divide the train series into a CV train series and a validation series, and find out which choice performs the best. Using this approach, we find that the best choice of $\phi$ is 0.25. Based on the relationship $\exp(-\phi d) \approx 0.05$, it implies that the correlation is negligible beyond a gap of 12 time points. 

Next, using the AIC values, it is found that an order of 1 suffices for the GLARMA method in this case. For the ACP model, following \cite{liboschik2015tscount} who analyzed the same data, we use autoregressive terms of order 1 and 12. They capture the short range serial dependence and the yearly seasonality, respectively. 

\begin{table}[!htb]
\centering
\caption{Parameter estimates (with standard errors) and goodness-of-fit measures from the three candidate models for the road accident data. }
\label{tab:estimates-accident}
\begin{tabular}{lcccc}
  \hline
Variable & GLM & GLARMA(1) & ACP(1,12) & Our model \\ 
  \hline
  Intercept & $2.812 (0.500)$ & $1.838 (0.517)$ & $2.881 (0.594)$ & $2.730 (0.736)$ \\ 
  Linear trend & $-0.004 (0.002)$ & $-0.004 (0.002)$ & $-0.004 (0.002)$ & $-0.004 (0.003)$  \\ 
  AR (lag 1) & - & $0.065 (0.079)$ & $0.062 (0.081)$ & - \\
  AR (lag 12) & - & - & $-0.072 (0.085)$ & - \\
  January & $-0.031 (0.113)$ & $-0.033 (0.11)$ & $-0.041 (0.114)$ & $-0.025 (0.139)$ \\ 
  February & $-0.426 (0.128)$ & $-0.427 (0.128)$ & $-0.461 (0.134)$ & $-0.428 (0.157)$ \\ 
  March & $-0.231 (0.127)$ & $-0.227 (0.128)$ & $-0.215 (0.133)$ & $-0.227 (0.156)$ \\ 
  April & $-0.269 (0.137)$ & $-0.265 (0.138)$ & $-0.272 (0.140)$ & $-0.259 (0.169)$ \\ 
  May & $-0.263 (0.166)$ & $-0.256 (0.169)$ & $-0.257 (0.169)$ & $-0.265 (0.207)$  \\ 
  June & $-0.096 (0.168)$ & $-0.089 (0.171)$ & $-0.08 (0.171)$ & $-0.078 (0.222)$ \\ 
  July & $-0.196 (0.211)$ & $-0.185 (0.215)$ & $-0.192 (0.212)$ & $-0.172 (0.279)$  \\ 
  August & $-0.218 (0.229)$ & $-0.207 (0.233)$ & $-0.212 (0.231)$ & $-0.175 (0.293)$ \\ 
  September & $-0.246 (0.170)$ & $-0.238 (0.172)$ & $-0.24 (0.173)$ & $-0.218 (0.206)$  \\ 
  October & $-0.008 (0.141)$ & $0 (0.142)$ & $0.009 (0.143)$ & $0.017 (0.172)$ \\ 
  November & $0.020 (0.113)$ & $0.023 (0.11)$ & $0.022 (0.113)$ & $0.022 (0.138)$ \\ 
  Kilometers & $0 (0.00004)$ & $0 (0.00004)$ & $0 (0.00004)$ & $0.001 (0.0001)$ \\ 
  Petrol price & $0.144 (2.368)$ & $0.079 (2.522)$ & $0.157 (2.364)$ & $1.373 (3.923)$ \\ 
  \hline
  Log-likelihood & $-441$ & $-441$ & $-441$ & $-313$  \\
  $R^2$ & $0.402$ & $0.404$ & $0.407$ & $0.710$ \\
  RMSE & $2.78$ & $2.78$ & $2.77$ & $1.94$  \\
   \hline
\end{tabular}
\end{table}

For all of the methods, an intercept term and an additive linear trend function are included in the mean structure. We also include monthly indicators to understand if the accidents differ significantly across different months. To avoid identifiability issues, the coefficient corresponding to December is taken as 0. The estimates and the standard errors of the parameters are presented in \Cref{tab:estimates-accident}. Values of the log-likelihood, $R^2$ and root mean square error (RMSE) on the train set are also given as goodness-of-fit measures in the same table.

So far as the estimates of the parameters are concerned, we notice that the results are more or less similar for all the models. February shows the maximum drop in the number of casualties. The coefficient for April also indicates a significant decrease in GLM, albeit the conclusions are the opposite in the other three models. In general, the summer months have negative impact on the number of accidents. Kilometers do not have an effect on the response variable while the coefficient for petrol price displays higher standard error in all of the models. The coefficient for the linear trend term is mildly negative everywhere, which depicts a decrease in the number of casualties as time progresses. We also note that the autoregressive terms are not significant in either of GLARMA or ACP. 

Turning attention to the goodness-of-fit measures, it is evident that our model outperforms the other two. Overall, the performance of GLM, GLARMA and ACP are comparable, whereas our model sees an approximately 75\% improvement in $R^2$. Log-likelihood and RMSE also improves by nearly 30\%.

Clearly, our specification works really well to capture the dependence pattern in the data. Refer to the model, \cref{eqn:full-model}. Our estimate for the error variance $\sigma^2$ is 0.029 whereas the estimate for $\sigma_w^2$ is 0.034. It shows that the zero-mean temporally correlated process explains marginally more variance than the random noise process, thereby establishing that the proposed specification is efficient for this type of count data analysis.

\begin{figure}[!ht]
\begin{center}
\includegraphics[width=0.8\textwidth,keepaspectratio]{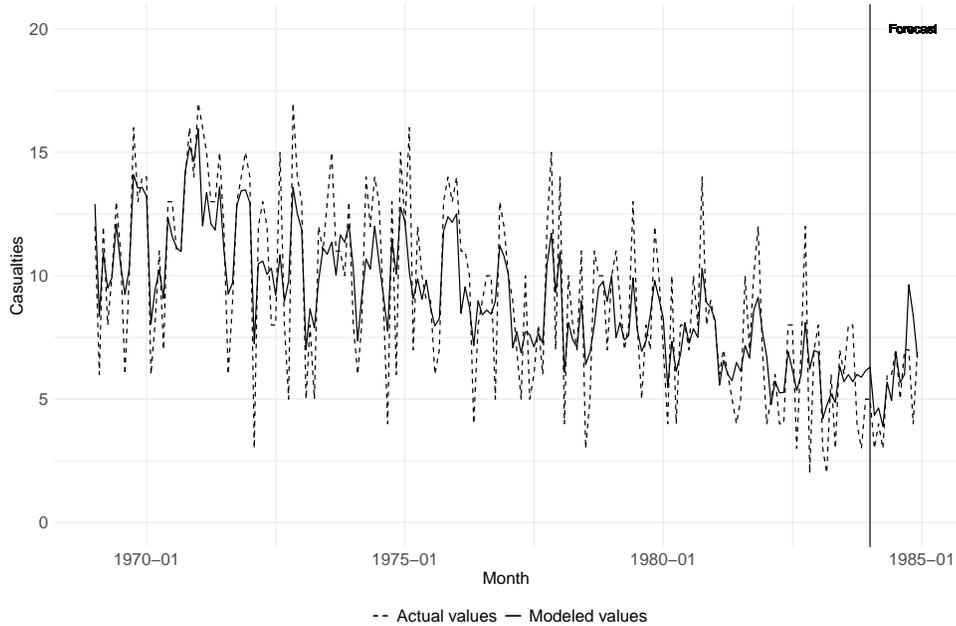}
\end{center}
\caption{Actual, fitted and predicted values for the number of casualties. The vertical line denotes the start of the forecast period.}
\label{fig:accident}
\end{figure}

Next, we take a look at the predictive accuracy. \Cref{fig:accident} shows the fitted and the predicted values for the entire series. The forecast period is of one year (12 observations) and it is evident that the point forecasts are quite close to the true realizations. Further, following \Cref{subsec:evaluation}, we compute the three scoring rules. \Cref{tab:scores-accident} shows these scores. 

\begin{table}[!htb]
\centering
\caption{Prediction scores for the competing models for the road accident data.}
\label{tab:scores-accident}
\begin{tabular}{lcccc}
  \hline
Scoring rule & GLM & GLARMA(1) & ACP(1,12) & Our model \\ 
  \hline
  Brier score & $0.1315$ & $-0.1599$ & $-0.1534$ & $-0.1707$ \\ 
  Spherical score & $-0.0930$ & $-0.4050$ & $-0.3956$ & $-0.4193$ \\ 
  Ranked probability score & $0.0003$ & $0.0001$ & $0.0001$ & $0.0001$ \\ 
   \hline
\end{tabular}
\end{table}

We see that both GLARMA and ACP perform considerably better than GLM in terms of all three scoring rules, with GLARMA being slightly better than ACP. Our model, meanwhile, has the best numbers in all three scoring rules. The scores show that the proposed approach is not only the best fitting model, but also has the highest predictive accuracy among all.

\subsection{Bike sharing data}
\label{subsec:bikes}

This dataset is from a bike sharing system named Capital Bike Sharing (CBS) at Washington, D.C., USA. \cite{fanaee2014event} analyzed this dataset before. It is interesting from the perspective of this paper for a few reasons. As it includes at least two full life-cycle of the system, similar to the previous application, it is suitable for both inferential and prediction purposes. For practical reasons, one would want to have good forecast for upcoming two to three months. Since the data is recorded on daily level, it provides us a good opportunity to understand the accuracy of the proposed approach when it comes to long-term forecasting. The response variable we will be working with is the total count of bike rentals, a combined number of both casual and registered rentals. \Cref{tab:summary-bikes} shows that the range of the count is 22 to 8714, with a mean of 4504. It is also highly variable, thereby providing us an opportunity to judge the performance of the proposed method for an over-dispersed data. Finally, the dataset includes information on weather conditions which are assumed to have an effect on the total count and the model can provide insight on that front as well. Summaries of these covariates are presented in \Cref{tab:summary-bikes}. The variable temperature denotes the normalized feeling temperature in Celsius. Humidity and windspeed values are normalized as well.

\begin{table}[!htb]
\caption{Summary of the response variable and the covariates for the bike sharing data.}
\label{tab:summary-bikes}
\centering
\begin{tabular}{lcccc}
  \hline
 & Range & Mean & Median & Standard deviation \\ 
  \hline
Count of bike rentals & $(22,8714)$ & 4504 & 4548 & 1937 \\ 
Temperature & $(0.079,0.841)$ & 0.474 & 0.487 & 0.163 \\ 
Humidity & $(0,0.972)$ & 0.628 & 0.627 & 0.142 \\ 
Windspeed & $(0.022,0.507)$ & 0.190 & 0.181 & 0.077 \\
   \hline
\end{tabular}
\end{table}

While implementing the four candidate models, along with the above three covariates, we include an intercept term, a linear trend and the monthly indicators (with coefficient of December as 0). Although the data is on a daily level, individual daily indicators do not improve the performance of the model. Instead, we add a binary indicator for working day as covariate, where 1 denotes a working day and 0 denotes a weekend or a holiday.

The CV scheme as before provides an estimate of $\phi$ equal to 0.1, which indicates that the temporal correlation is not significant beyond a month. For GLARMA, an autoregressive order of 1 turns out to be the most appropriate specification whereas for ACP, we use orders of 1 and 7, in view of that it is a daily dataset.

The estimates of the coefficients are presented in \Cref{tab:estimates-bikes}. Unlike the road accident dataset, the results are somewhat different for the four models in this case. The linear trend term is positive here, depicting a steady increase in the count of rentals over time. The autoregressive terms in GLARMA and ACP are significant and show that the numbers are positively correlated for lags 1 and 7. So far as environmental factors are considered, higher temperature, lower humidity and lower windspeed see an increase in the total number of bike rentals. These coefficients and the standard errors for all four models are in the same range. The same is reflected in the coefficients of the month variables as well, where we observe that the effects are higher for summer and autumn months and are lower for rainy and winter months. Finally, the number of bike rentals tend to be higher for the working days in case of all the models. All of these results are expected, as better weather should cause a higher number of casual rentals while usual weekday chores tend to increase the number of registered rentals.

\begin{table}[!htb]
\centering
\caption{Parameter estimates (with standard errors) and goodness-of-fit measures from the three candidate models for the bike sharing data.}
\label{tab:estimates-bikes}
\begin{tabular}{lcccc}
  \hline
Variable & GLM & GLARMA(1) & ACP(1,7) & Our model  \\ 
  \hline
  Intercept & $7.661 (0.006)$ & $6.724 (0.009)$ & $5.322 (0.024)$ & $7.685 (0.152)$ \\ 
  Linear trend & $0.001 (0)$ & $0.001 (0)$ & $0.001 (0)$ & $0.001 (0)$ \\ 
  AR (lag 1) & - & $0.494 (0.003)$ & $0.261 (0.003)$ & - \\
  AR (lag 7) & - & - & $0.061 (0.002)$ & - \\
  January & $-0.01 (0.004)$ & $-0.036 (0.008)$ & $-0.003 (0.004)$ & $-0.023 (0.132)$ \\ 
  February & $0.074 (0.004)$ & $0.038 (0.008)$ & $0.038 (0.004)$ & $0.092 (0.145)$ \\ 
  March & $0.285 (0.004)$ & $0.288 (0.008)$ & $0.199 (0.004)$ & $0.225 (0.147)$ \\ 
  April & $0.364 (0.004)$ & $0.357 (0.007)$ & $0.244 (0.004)$ & $0.336 (0.149)$ \\ 
  May & $0.416 (0.004)$ & $0.419 (0.008)$ & $0.276 (0.005)$ & $0.353 (0.154)$ \\ 
  June & $0.306 (0.005)$ & $0.305 (0.008)$ & $0.18 (0.005)$ & $0.269 (0.155)$ \\ 
  July & $0.157 (0.005)$ & $0.166 (0.008)$ & $0.078 (0.005)$ & $0.133 (0.158)$ \\ 
  August & $0.221 (0.005)$ & $0.248 (0.008)$ & $0.131 (0.005)$ & $0.178 (0.154)$ \\ 
  September & $0.32 (0.004)$ & $0.341 (0.008)$ & $0.215 (0.005)$ & $0.263 (0.151)$ \\ 
  October & $0.301 (0.005)$ & $0.298 (0.008)$ & $0.192 (0.005)$ & $0.121 (0.148)$ \\ 
  November & $0.179 (0.005)$ & $0.196 (0.008)$ & $0.122 (0.005)$ & $0.154 (0.126)$ \\ 
  Working day & $0.025 (0.001)$ & $0.019 (0.001)$ & $0.029 (0.001)$ & $0.054 (0.019)$ \\ 
  Temperature & $1.124 (0.009)$ & $1.085 (0.011)$ & $0.805 (0.01)$ & $1.37 (0.161)$ \\ 
  Humidity & $-0.66 (0.005)$ & $-0.795 (0.005)$ & $-0.607 (0.005)$ & $-0.953 (0.068)$ \\ 
  Windspeed & $-0.788 (0.009)$ & $-0.556 (0.009)$ & $-0.695 (0.009)$ & $-0.974 (0.13)$ \\
  \hline
  Log-likelihood & $-58441$ & $<-10^6$ & $-52064$ & $-3202$ \\
  $R^2$ & $0.812$ & $0.858$ & $0.839$ & $0.998$ \\
  RMSE & $832$ & $724$ & $770$ & $5.37$  \\
   \hline
\end{tabular}
\end{table}

\begin{figure}[!ht]
\begin{center}
\includegraphics[width=0.8\textwidth]{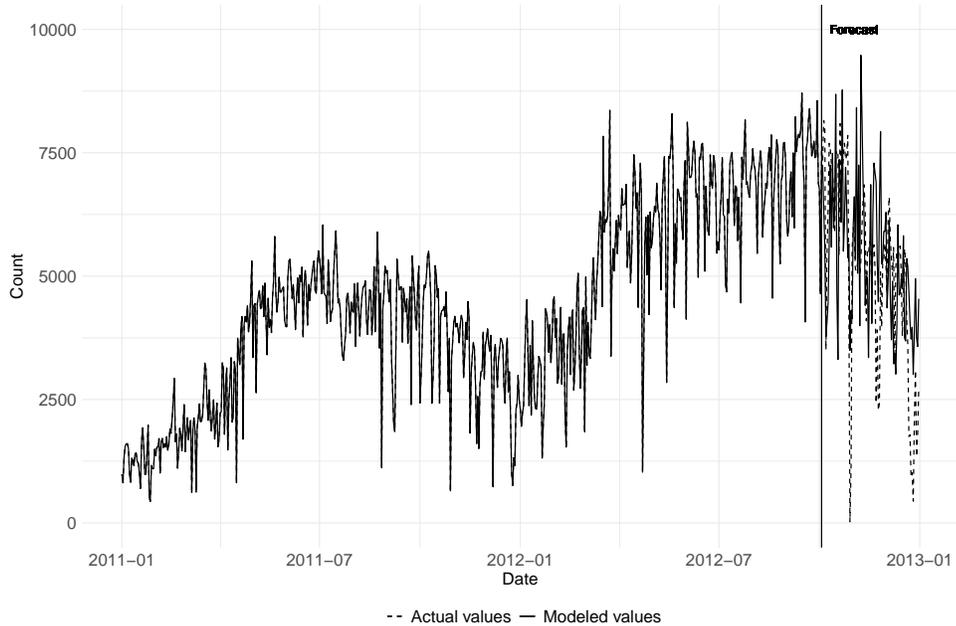}
\end{center}
\caption{Actual, fitted and predicted values for the bikes data. The vertical line denotes the start of the forecast period.}
\label{fig:bikes}
\end{figure}

From the goodness-of-fit measures, we see that the fit is much better for our model. In fact, the plot in \Cref{fig:bikes} depicts that the fitted values are aligning well with the observed values throughout the training period. The same plot also shows that the predictions are very good for the 90-day forecast period. Evidently, the exponentially decaying covariance function captures the temporal dependence pattern well. The estimated variance of the temporally dependent process ($\sigma_w^2$) is 0.042, once again higher than the variance of the white noise process ($\sigma^2$) which was estimated as 0.037. That our model is better than the autoregressive specifications of GLARMA or ACP is further confirmed by \Cref{tab:scores-bikes}. There, we see that the quadratic score or the spherical scores are much lower for our model. We point out that the ACP model is expected to perform well for an over-dispersed data, but our model beats it both in terms of fitting and predicting. We also experimented with a autoregressive conditional negative binomial model, which differs from ACP by a negative binomial specification in lieu of Poisson. The performance did not improve there and it is worse than the proposed method as well. Finally, note that the generalized linear model works much worse than the other three models in terms of the prediction scores. It is natural in view of the fact the other models capture the dependence nature of the data in one way or the other.

\begin{table}[!htb]
\centering
\caption{Prediction scores for the competing models for the bike-sharing data.}
\label{tab:scores-bikes}
\begin{tabular}{lcccc}
  \hline
Scoring rule & GLM & GLARMA(1) & ACP(1,7) & Our model \\ 
  \hline
  Brier score & $0.0971$ & 0.0028 & 0.0031 & $-0.00004$ \\ 
  Spherical score & $-0.00002$ & $-0.0056$ & $-0.0064$ & $-0.0102$ \\ 
  Ranked probability score & $0.5143$ & $0.0996$ & $0.0950$ & $0.1123$ \\
   \hline
\end{tabular}
\end{table}

\section{Concluding remarks}
\label{sec:remarks}

To summarize, in this study, we have developed a new method to analyze and forecast time series count data. The simulation study and the applications show that the method works well across various scenarios. It is much better than the other popular models both in terms of explaining the variation in the data and in terms of predictive accuracy. It is also precise, easy to interpret and flexible enough to include more regressors in the mean structure as necessary. The method is applicable for irregularly spaced time points, and hence can be used efficiently for such datasets. 

We notice that the proposed approach works much better than the other competing methods both when the values are less (example is the road accident data) and when the numbers and the variability are higher (example is the bike sharing data). As a future direction to this work, it can be interesting to develop a similar model for zero-inflated Poisson regression setup. One can also extend the model to negative binomial and zero-inflated negative binomial variables as well. That would allow us to deal with count data in various capacities across different disciplines.

Another potential future scope of this work is related to the specification of the covariance structure. We have used a special class of Mat\'ern covariance function in the dependent process. One can think of relaxing that assumption and consider a more general class of time-dependent processes. 

Further, recall that the optimal value of the exponential decay parameter $\phi$ is estimated using a cross-validation scheme. While this helps in reducing the computational burden, a complete Bayesian approach may provide more precise estimate. One can also work out the classical maximum likelihood type estimators for the model.

\section*{Funding source}

This research did not receive any specific grant from funding agencies in the public, commercial, or not-for-profit sectors.

\section*{Declarations of interest}

The author declares no conflict of interest.

\bibliography{mybibfile}

\end{document}